\documentclass[aps,pra,preprint,superscriptaddress,amsmath,amssymb]{revtex4-2}

\usepackage{graphicx}% Include figure files
\usepackage{dcolumn}% Align table columns on decimal point
\usepackage{bm}
\usepackage[utf8]{inputenc}
\usepackage[T1]{fontenc}
\usepackage{booktabs, array, mathptmx, float, tabularx, booktabs, lipsum, amsmath,multirow}
\usepackage{siunitx, xcolor}
\usepackage[version=4]{mhchem}
\usepackage{color}
\usepackage{setspace}
\usepackage{subfigure}
\usepackage[colorlinks,linkcolor=blue,anchorcolor=blue,citecolor=blue]{hyperref}

\begin{document}
\begin{spacing}{2.0}
\title{Far-field model of two-color laser driven  terahertz radiation including field element interference and plasma response}

\affiliation{Department of Physics and Beijing Key Laboratory of Opto-electronic Functional Materials and Micro-nano Devices, Renmin University of China, Beijing 100872, China}
\affiliation{Key Laboratory of Quantum State Construction and Manipulation (Ministry of Education), Renmin University of China, Beijing, 100872, China}
\affiliation{IFSA Collaborative Innovation Center, Shanghai Jiao Tong University, Shanghai 200240, China}

\author{Nan Li}
\affiliation{Department of Physics and Beijing Key Laboratory of Opto-electronic Functional Materials and Micro-nano Devices, Renmin University of China, Beijing 100872, China}
\affiliation{Key Laboratory of Quantum State Construction and Manipulation (Ministry of Education), Renmin University of China, Beijing, 100872, China}

\author{Wei-Min Wang}
\email{weiminwang1@ruc.edu.cn}
\affiliation{Department of Physics and Beijing Key Laboratory of Opto-electronic Functional Materials and Micro-nano Devices, Renmin University of China, Beijing 100872, China}
\affiliation{Key Laboratory of Quantum State Construction and Manipulation (Ministry of Education), Renmin University of China, Beijing, 100872, China}
\affiliation{IFSA Collaborative Innovation Center, Shanghai Jiao Tong University, Shanghai 200240, China}

\begin{abstract}
The two-color laser field scheme for broad-band terahertz (THz) radiation from air has been intensively and broadly investigated due to the simplicity in technology and relative high yield efficiency. Experiments showed that the detected THz angular distribution is usually conical with a dip in the laser propagation axis which is unfavorable for its application, and the formation of an oscillating tail in the THz waveform has not yet been well understood. Here, we develop an electric field element interference model in which each local current source has a phase velocity determined by the laser pulse and a Gaussian temporal profile, rather than an ideal point source, as well as the current oscillation caused by the plasma dynamics is considered.  By introducing the temporal interference, effects of the phase velocity and the Gaussian profile of the current can correct the THz angular distribution. Therefore, by adjusting the laser pulse duration and the group velocity, the THz angular distribution can be controlled and the dip could be even eliminated. Besides, the plasma dynamics can result in the formation of an oscillating tail in the THz waveform and sunken spikes in the corresponding spectrum. This is similar to the experimental results which are usually explained by the water vapor absorption. In turn, the spikes in the spectrum could provide a way to measure the plasma filament density. 
\end{abstract}
\maketitle

\section{Introduction}
Terahertz (THz) waves refer to electromagnetic waves with the frequencies at the range of $10^{11}$ Hz-$10^{13}$ Hz, corresponding to the wavelength range of 3 mm-0.03 mm, which are located on the transition region between microwaves and infrared waves. Because of the unique frequency range, THz waves have extensive applications in medical diagnoses, security inspections and imaging detections \cite{pickwell2006biomedical,shen2005detection,gaal2006nonlinear,jiang1998single,dobroiu2006terahertz,tonouchi2007cutting,zhang2017extreme,moldosanov2017terahertz,massaouti2013detection}. Therefore, THz radiation source generations and THz detections via different approaches are a hot issue for a long time and they have made great progresses in recent decades \cite{auston1984cherenkov,bass1962optical,nahata1996wideband,scalari2006electrically,ito2005continuous,gallot1999electro,tani2000detection,koulouklidis2020observation,oh2014generation,johnson2013thz,kuk2016plasma,kim2014terahertz,oh2013intense}. With the development of the ultrashort ultra-intense laser technology, the ultra-intense femtosecond laser pulse has been widely applied for broad-band strong THz radiation generations via the formed plasma from gases \cite{cook2000intense,bartel2005generation,zhong2006terahertz,kim2008coherent,kuk2016generation,babushkin2010ultrafast}, solids \cite{liao2016demonstration,liao2019multimillijoule}, and liquids \cite{dey2017highly,jin2018terahertz,zhang2019strong,chen2022scaling}. Among them, the THz generation from air irradiated by the two-color laser pulse \cite{cook2000intense,bartel2005generation,zhong2006terahertz,kim2008coherent,kuk2016generation,babushkin2010ultrafast} has been generally investigated because of its simplicity and stability in experiments and relatively high yields. 

The THz mechanism was firstly attributed to a four-wave mixing process in the gas plasma, which can explain partial characteristics of the THz radiation such as the THz polarization and the dependence on the relative phase of the two-color laser field \cite{kress2004terahertz,xie2006coherent}. Later, Kim et al. proposed that the origin of the THz radiation is a net current generated by the symmetry-broken laser field ionization of gases \cite{kim2007terahertz, wang2008strong, kim2009generation}, and developed a far-field model successfully explaining the THz conical emission and the intensity dependency \cite{you2012off,gorodetsky2014physics}, but could not give the THz waveform and the THz frequency self-consistently. To include the contribution of the plasma dynamics in the THz generation, Wang et.al \cite{wang2008strong,wang2011towards,wang2015tunable} focused on the evolution of the net current in plasmas and proposed a near-field plasma model, which can obtain the near single-cycle waveform and the THz central frequency determined by the plasma oscillation frequency. However, it is hard to verify this model because of the difficulty of measuring the plasma filament density experimentally.

In this article, we combine the two models above together and extend the THz far-field description with the formation of the net current and its evolution in plasmas. In the previous models \cite{you2012off,gorodetsky2014physics}, each local current source is considered as an ideal point source and emits THz radiation with the frequency determined by the local plasma density. We consider that each local current source has a phase velocity determined by the group velocity of the laser pulse and the source has a Gaussian temporal profile given by the laser duration and the plasma collision. Then, both the spatial and the temporal interference of the emitted THz radiation from each current source can be included.  Since the net current has a phase velocity and a temporal profile, we find that the dephasing effect can be influenced by the current duration (or the laser duration) and the laser group velocity which could be adjusted by the laser wavelength or the plasma filament density. By decreasing the current/laser duration and the laser group velocity, the conical THz angular distribution with a dip in the laser propagation axis usually observed in experiments could be eliminated. 

When the plasma dynamics is considered, an oscillating current appears following the Gaussian current pulse, which can bring an oscillating tail behind the mian THz pulse. The oscillating tail introduces sunken spikes at the corresponding plasma frequencies, which look similar to the absorption peaks. Such a long-time weak oscillating tail in the THz waveform and sunken spikes in the spectrum were observed in experiments \cite{kim2007terahertz,liu2016enhanced,tan2022water}, which are usually interpreted as the absorption of the THz radiation by water vapor. However, the waveform of this oscillation tail is very different from the THz main pulse. The strength of the tail suddenly decreases by one order of magnitude from the main pulse and the tail maintains an oscillation for few picoseconds while the main pulse last for only nearly one cycle. Note that the experimental comparison with and without vapor absorption shows \cite{liu2010broadband} the spikes in the THz spectrum are very densely distributed in the frequency range of 1 THz to 7 THz. However, the THz spectrum observed from two-color laser experiments \cite{liu2016enhanced,kim2007terahertz} has several isolated spikes. These features can not be completely explained by the absorption of water vapor, but could be attributed to the evolution of the dynamic current in the plasma in our model. Furthermore, the oscillating tail and the corresponding spikes in the spectrum could provide a way to measure the plasma filament density. Each spike in the spectrum corresponds to a plasma frequency $\omega_{spike}=\frac{1}{\sqrt{2}}\omega_{p}=\frac{1}{\sqrt{2}}\sqrt{\frac{n_{e0}e^{2}}{m\epsilon_{0}}}$, which is connected with the plasma density $n_{e0}$. Thus, one can read plasma density information in the spectrum from these spikes.

\section{Model and simulation results}

\subsection{The far-field model with a Gaussian profile of the plasma current}

In the two-color field scheme, the THz radiation is generated from a dynamic current as other electromagnetic emission in the classical electrodynamics \cite{jackson1999classical}. This current originates from a gas ionization process and then follows the physics of the current evolution in plasmas. The photoionization current model \cite{kim2007terahertz} shows that the symmetry-broken electric field generates a net current which leads to the THz radiation. We will start our investigation from this idea and extend the spatial interference of the point sources (spherical wave) to the temporal interference of the current sources with a Gaussian temporal profile. Then, the angular distribution and the frequency spectrum of the THz radiation can be self-consistently calculated, rather than the assumption of monochromatic radiation \cite{you2012off}. Besides, we consider the current source with a phase velocity equal to the laser group velocity to correct the spatial and the temporal interference results. Finally, the plasma oscillation is also considered, which can explain the oscillating tail of the THz radiation observed in experiments.

\begin{figure}[!htbp]
    \centering
     \includegraphics[width=8.6cm]{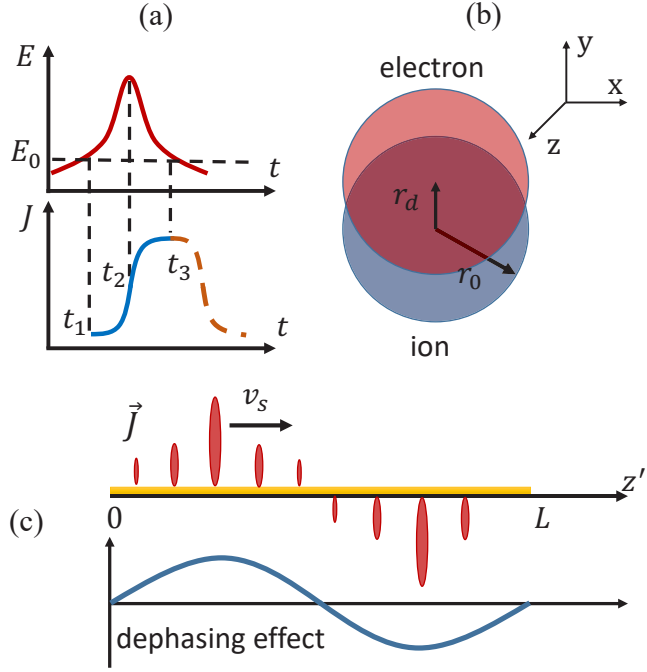}
	\begin{spacing}{1.2}
	\caption{$\left(\mathrm{a}\right)$The Gaussian laser pulse (up) and the corresponding current (down), where $E_{0}$ is the ionization threshold and $t_{1}$, $t_{2}$ and $t_{3}$ are three different times. $\left(\mathrm{b}\right)$The charge oscillation caused by the separation of the electron and the ion density center in the laser polarization direction ($y$ direction), where $r_{d}$ is the deviation of the electron center from the ion center, $r_{0}$ is the plasma filament radius, and $r_0 \gg r_{d}$. $\left(\mathrm{c}\right)$The dynamic current (up) including the dephasing effect (down), where $v_{s}$ is the phase velocity of the current, $L$ the plasma filament length, and $\vec{J}$ is a temporal profile of the current at a given $z^{'}$. During the propagation, the current magnitude and the polarity change along with $z^{'}$ because of the dephasing effect.}
	\end{spacing}
	\label{fig1}
\end{figure}

We consider a gas molecule at the position $\vec{x}^{\prime}$. When a two-color laser pulse passes through it, the molecule is ionized at the time $t^{\prime}$ and an electron is born with an initial velocity of zero. In the laser field, it gains a velocity:
\begin{equation}\label{eq1}
	v\left(t\right)=-\frac{e}{m_{e}}\int_{t^{\prime}}^{t}E_{L}\left(\tau\right)\mathrm{d}\tau,
\end{equation}
where $e$ and $m_{e}$ are separately the electron charge and mass. A two-color laser electric field can be written as:
\begin{equation}\label{eq2}
	E_{L}\left(r,t\right)=E_{\omega}f_{1}e^{-\frac{t^{2}}{\tau_{1}^{2}}}\cos\left(\omega t\right)+
	E_{2\omega}f_{2}e^{-\frac{t^{2}}{\tau_{2}^{2}}}\cos\left(\omega t+\theta\right),
\end{equation}
where $f_{1}\left(r\right)$ and $f_{2}\left(r\right)$ are transverse profiles of the fundamental wave laser field and its second harmonic, respectively, $\tau_{1}$ and $\tau_{2}$ are pulse durations, $\theta$ is the relative phase between the fundamental wave and its second-harmonic. We consider that the laser spot size and the duration are far greater than the laser wavelength and the period, so its transverse and longitudinal distributions can be ignored. Then, the laser field can be simplified as:
\begin{equation}\label{eq3}
	E_{L}\left(t\right)=E_{\omega}\cos\left(\omega t\right)+E_{2\omega}\cos\left(\omega t+\theta\right).
\end{equation}
The electron velocity can be expressed as:
\begin{equation}\label{eq4}
    v\left(t\right)=-\frac{eE_{\omega}}{m_{e}\omega}\sin\left(\omega t\right)-\frac{eE_{2\omega}}{2m_{e}\omega}\sin\left(\omega t+\theta\right)+v_{d}\left(t^{\prime}\right),
\end{equation}
where
\begin{equation}\label{eq5}
	v_{d}\left(t^{\prime}\right)=\frac{eE_{\omega}}{m_{e}\omega}\sin\left(\omega t^{\prime}\right)+\frac{eE_{2\omega}}{2m_{e}\omega}\sin\left(\omega t^{\prime}+\theta\right).
\end{equation}
Eqs. \textcolor{blue}{(}\ref{eq4}\textcolor{blue}{)} and \textcolor{blue}{(}\ref{eq5}\textcolor{blue}{)} show that the generated electron has an oscillation velocity and a drift velocity $v_{d}\left(t^{\prime}\right)$, where the drift velocity is determined by the laser field strength and the relative phase at the ionization position. Supposing the electron density is $dN_{e}\left(t^{\prime}\right)$, we obtain a current $dN_{e}\left(t^{\prime}\right)v\left(t\right)$. Note that the current or the velocity oscillating at the laser frequency does not generate THz-band radiation \cite{roskos2007broadband} and what really contributes to is the drift current $dN_{e}\left(t^{\prime}\right)v_{d}\left(t^{\prime}\right)$. With a suitable relative phase $\theta$,a stronger laser field causes a higher ionization probability, a larger drift velocity, and a stronger net current eventually. It should be noticed that the relative phase between the fundamental wave and its second harmonic changes periodically in the propagation because they have different phase velocities. A dephasing length $L_{dp}$ can be defined as:
\begin{equation}\label{add1}
	L_{dp}=\frac{\lambda_{2\omega}}{n_{2\omega}-n_{\omega}},
\end{equation}
where $n_{\omega}$ and $n_{2\omega}$ are refractive indices for the fundamental wave and its second harmonic, respectively, and $\lambda_{2\omega}$ is the wavelength of the second harmonic.

With the two-color laser pulse passing through a local position, the current is continuously accumulated, as shown in Fig. 1(a). For a Gaussian laser pulse, at the time point $t_{1}$, the current accumulation speed (blue line slope) is slow due to the low laser intensity; at the time point $t_{2}$, the laser intensity reaches the peak and the current increases fastest; at the time point $t_{3}$ when the laser is about to leave, the current increases slowly again and reaches a peak value.  The rising edge of the current from $t_{1}$ to $t_{3}$ have a timescale approaching to the laser duration. Then, the current is attenuated by the collision \cite{sprangle2004ultrashort} of the plasma composed of electrons and nearly immobile ions. Hence, a typical lifetime of the current is several hundred of femtoseconds, which is determined by the duration of the laser pulse plus the characteristic time of the plasma collision \cite{sprangle2004ultrashort}. The plasma collision determines the current falling edge with a timescale of tens of or hundreds of femtoseconds. For simplicity, we take a Gaussian time profile of the current with the falling edge equal to the rising edge, since the timescale of the two edges is at the same order.

As the laser pulse propagates forward, new net currents are continuously excited at new positions. Actually, the net current does not move with the laser pulse since the electron velocity is transverse, and therefore, the formed currents have a phase velocity $v_{s}$ equal to the laser group velocity. The formed currents only oscillate transversely at local positions, as displayed in Fig. 1(b). These currents formed at different positions follow a similar evolution process, but the overall magnitude and the polarity are different because of the dephasing effect as shown in Fig. 1(c). The dynamic transverse current can emit the THz radiation with vector potential $\vec{A}$ in the far field:
\begin{equation}\label{eq6}
\vec{A}\left(\vec{x},t\right)=\frac{\mu_{0}}{4\pi}\int \mathrm{d}V^{\prime}\frac{\vec{J}\left(\vec{x}^{\prime},t-\frac{r}{c}\right)}{r},
\end{equation}
we can further obtain the THz electric field $\vec{E}$:
\begin{equation}\label{eq7}
	\vec{E}\left(\vec{x},t\right)=-\frac{\partial \vec{A}\left(\vec{x},t\right)}{\partial t},
\end{equation}
and the THz angular distribution:
\begin{equation}\label{eq8}
    I\left(\mit\Theta\right)=\int_{-\infty}^{\infty}\left|E\left(\mit\Theta,t\right)\right|^{2}\mathrm{d}t.
\end{equation}
We consider the current has a temporal profile of Gaussian with a duration of $\tau$, i.e.,

\begin{equation}
	\label{eq9}
	\vec{J}\left(\vec{x}^{\prime},t\right)=\vec{J}_{0}\cos\left(\frac{2\pi}{L_{dp}}z^{\prime}+\phi_{0}\right)\tau^{\frac{1}{2}}e^{-\left(\frac{t-\frac{z^{\prime}}{v_{s}}}{\tau}\right)^2},
\end{equation}

where the cosine term represents the modulation of the current due to dephasing effect, $L_{dp}$ is the dephasing length which comes from the phase velocity difference between the fundamental wave and its second harmonic, $\phi_{0}$ is their initial relative phase at the beginning edge of the plasma filament, and $\tau$ is the current duration.

In our model (also in \cite{you2012off,gorodetsky2014physics}), the field ionization is not included and the net current via the ionization is directly taken. For the two-color scheme, the THz radiation is generated from the net current and this current is a specific result of the symmetry-broken two-color laser field. The relative phase between the fundamental wave and its second harmonic changes with the propagation of the two-color laser. This relative phase determines the magnitude and the polarity of the current. If a single-color laser is used, the symmetry of the laser field ionization cannot be broken and a net current is not generated. Therefore, our model cannot be applied in the single-color scheme \cite{sprangle2004ultrashort,d2007conical}.

\begin{figure}[!htbp]
	\centering
     \includegraphics[width=8.6cm]{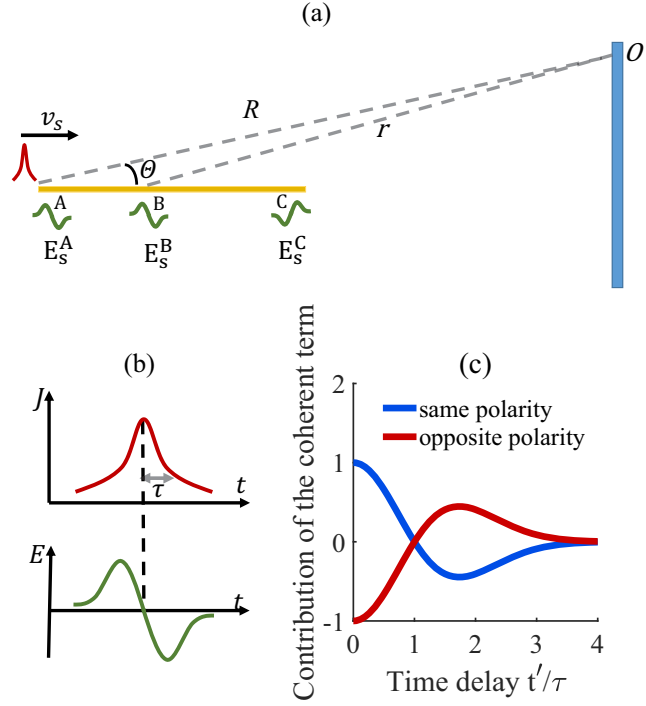}
	\begin{spacing}{1.2}
	\caption{The schematic diagram of the electric field element interference model. $\left(\mathrm{a}\right)$ $E_{S}^{A}$,$E_{S}^{B}$ and $E_{S}^{C}$ are electric field elements excited by the currents at three positions $A$, $B$ and $C$, where  $A$ is the beginning point of the filament and $C$ is the ending point, $R$ is the distance from $A$ to the far field point $O$, $\mit\Theta$ is the included angle between $OA$ and the filament. $\left(\mathrm{b}\right)$ An electric field element emitted from a Gaussian current with a characteristic duration of $\tau$ at a given position. $\left(\mathrm{c}\right)$ The contribution of a coherent term (in Eq. \textcolor{blue}{(}\ref{eq11}\textcolor{blue}{)}) of two electric field elements to the THz intensity, where the time delay between the two elements is $t^{\prime}$. Note that positive and negative values denote positive and negative contributions to the THz intensity observed at the far-field point $O$.\label{fig2}}
  \end{spacing}
\end{figure}

Next, we analyze the mechanism of the dephasing effect on the THz angular distribution. As the laser pulse propagates forward, the plasma filament and currents are formed. For instance, when the pulse passes through positions $A$, $B$ and $C$ marked in Fig. 2(a), the THz radiation sources are emitted with the electric field elements $E_{S}^{A}$, $E_{S}^{B}$ and $E_{S}^{C}$, respectively. These electric field elements propagate to the far-field point $O$ and superpose temporally to form a THz pulse. We assume that the plasma filament is composed of $N\left(N\rightarrow\infty\right)$ slices with thickness $L/N$, where $L$ is the length of the plasma filament. Then, each slice contains a current pulse emitting an electric field element. The final THz pulse at the far-field point $O$ is the coherent superposition of $N$ electric field elements. For a Gaussian current with a characteristic duration $\tau$ as shown in Fig. 2(b), a generated electric field element can be expressed as:
\begin{equation}\label{eq10}
	E^{i}_{s}(t)=(\pm 1)_{i} \times A_{s}^{i}\frac{2(t-t_{i})}{\tau^{\frac{3}{2}}}e^{-(\frac{t-t_{i}}{\tau})^{2}},
\end{equation}
where $E_s^i(t)$ indicates this contribution comes from the i-th slice, $(\pm 1)_{i}$ represents the polarity of the electric field element, $A_{s}^{i}$ is the amplitude of this field element when it arrives at the far-field point $O$, and $t_{i}$ is the time delay. Then the THz intensity is given by: 
\begin{equation}\label{eq11}
  I \propto \sum_{i<j}^{N}\sum_{j=1}^{N}\left[\int_{-\infty}^{\infty}2E_{s}^{i}\left(t\right)E_{s}^{j}\left(t\right)\mathrm{d}t\right]+\sum_{i=1}^{N}\int_{-\infty}^{\infty}\left[E_{s}^{i}\left(t\right)\right]^{2}\mathrm{d}t,
\end{equation}
for further analysis, we give its integral result:
\begin{equation}\label{eq12}
	I \propto \sum_{i<j}^{N}\sum_{j=1}^{N}(\pm1)_{i}(\pm1)_{j}2A_{s}^{i}A_{s}^{j}e^{-\frac{1}{2}(\frac{t_{i}-t_{j}}{\tau})^{2}}[1-(\frac{t_{i}-t_{j}}{\tau})^{2}]+\sum_{i=1}^{N}{A_{s}^{i}}^{2},
\end{equation}
the first term in Eq. \textcolor{blue}{(}\ref{eq11}\textcolor{blue}{)} is the coherent superposition of $N(N-1)/2$ pairs of electric field elements and the second one is a constant. When the laser pulse reaches the slice $A$ at the time $0$, the corresponding electric field element at the point $O$ starts to appear at the time $R/c$. When the laser pulse moves to the slice $B$ at the time $t$, the corresponding electric field element at the point $O$ starts to appear at the time $R/c+(1-\frac{v_{s}}{c}cos\Theta)t$. Considering the interference of two electric field elements with a time delay $t^{\prime}$, we have: 
\begin{equation}\label{eq13}
	t^{\prime}=(1-\frac{v_{s}}{c}cos\Theta)\frac{\Delta L}{v_{s}},
\end{equation}
where $\Delta L$ is the distance between the two slices. As illustrated in Fig. 2(c), with $t^{\prime}=0$, the coherent term will have a positive contribution to the total THz intensity if the two elements have the same polarity (blue line). Otherwise, the coherent term will have a negative contribution and weaken the total THz intensity if the two elements have opposite polarity (red line). While $t^{\prime}>4\tau$, the two elements are almost separated temporally and the coherent term have no contribution to the total THz intensity. 

The ratio $t^{\prime}/\tau$ reflects the dephasing effect and influences the THz angular distribution strongly. For example, we take typical parameters as: $v_{s}=0.9997c$ (the group velocity of the 800 nm laser in air), the filament length $L=30$ mm, and the current duration $\tau=0.15$ ps. When the observation point is at $0^{\circ}$, the time delay $t^{\prime}$ between $A$ and $C$ is 0.03 ps, far less than $\tau$. While the observation point is at $5^{\circ}$, $t^{\prime}$ rises to 0.41 ps, closing to $3\tau$. Therefore, the observed THz angular distribution can be affected much stronger at $\Theta=0^{\circ}$ than $\Theta=5^{\circ}$ due to the dephasing effect. If there is no dephasing effect and all electric field elements have the same polarity, all coherent terms make positive contributions and lead to a THz intensity peak appearing at $0^{\circ}$, since $t^{\prime}$ is much smaller than $\tau$. For the angle of $5^{\circ}$ with $t^{\prime}\approx3\tau$, this large time delay makes the coherent terms have nearly no contributions to the THz intensity, which causes the intensity weaker than that at $0^{\circ}$. Therefore, the THz angular distribution is a single peak without the dephasing effect considered, as shown in \cite{you2012off}. However, With the dephasing effect, the angular distribution will transit from a single peak to double peaks. This is because any coherent pair with opposite polarity will decrease the THz intensity at small angles. For example, in a typical condition, the number of positive and negative currents is half in half, the numbers of coherent pairs with the same polarity and the opposite polarity are $\frac{N}{2}(\frac{N}{2}-1)$ and $\frac{N}{2}\frac{N}{2}$, respectively, the latter is $\frac{N}{2}$ more than the former. For $0^{\circ}$, the value $t^{\prime}/\tau$ is close to zero for almost all coherent pairs, therefore, the $\frac{N}{2}$ net negative contributions make the THz intensity decrease at this angle. These negative contributions can finally make the THz intensity at $0^{\circ}$ much smaller than other medium angles. These will be illustrated by our simulation in Figs. \ref{fig3} and \ref{fig4}.

\subsection{Control of the THz angular distribution by reducing the dephasing effect}

\begin{figure}
     \centering
     \includegraphics[width=8.6cm]{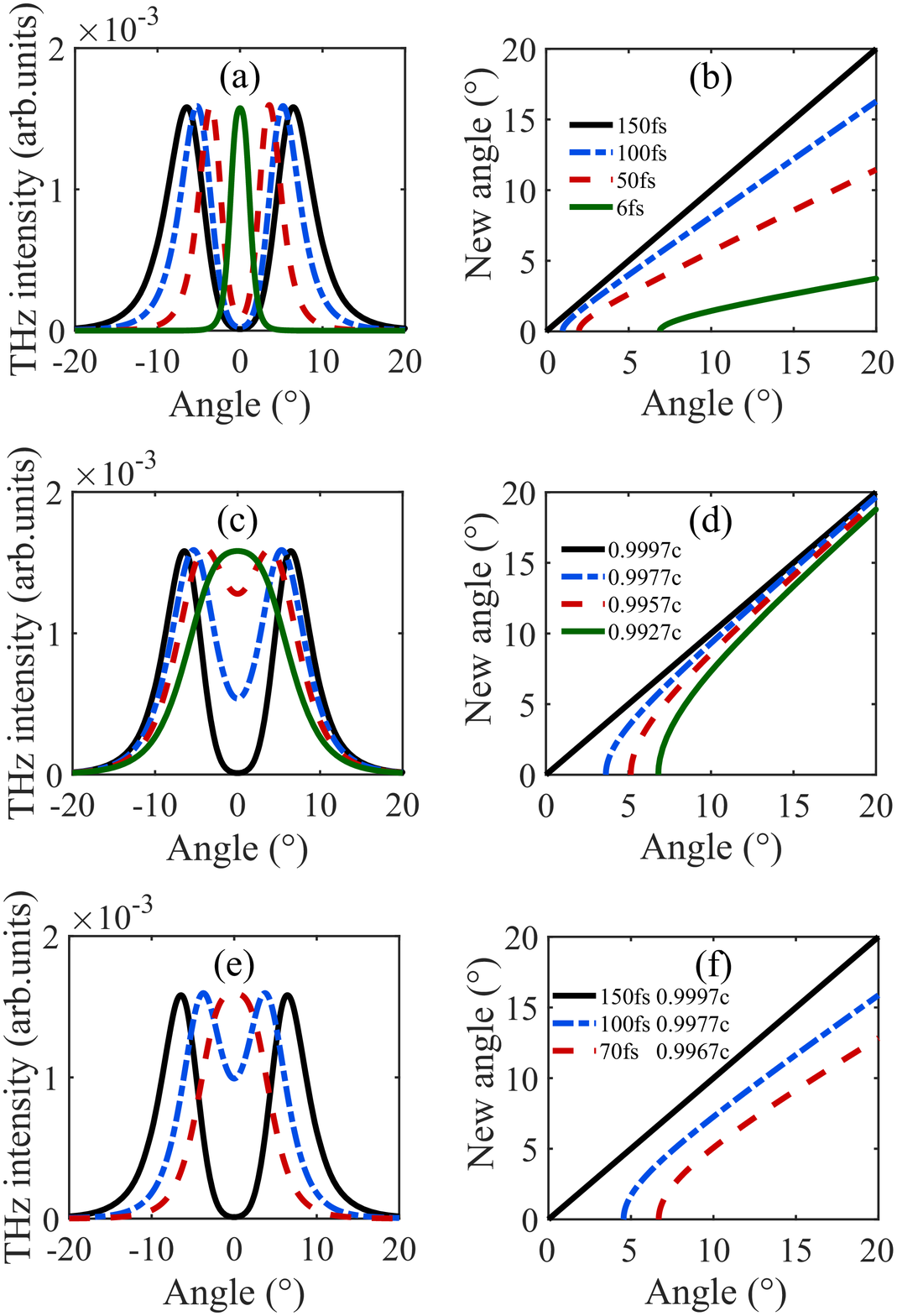}
\begin{spacing}{1.2}	
	\caption{THz angular distributions with (a) different current pulse durations $\tau$ (c) different laser group velocities $v_{s}$, and (e) different $\tau$ and $v_{s}$. [(b), (d) and (f)] The new angle as a function of the original angle when we change the typical parameters($v_{s}=0.9997c$, $\tau=$150 fs) to another set of parameters corresponding to [(a), (c) and (e)], respectively. Here the initial laser relative phase $\phi_{0}=0.5\pi$ and filament length $L=L_{dp}=$20 mm are taken in all subfigures, $v_{s}=0.9997c$ is fixed in (a) and (b), and $\tau=$150 fs is fixed in (c) and (d).\label{fig3}}
\end{spacing}
\end{figure}

In the previous model, when the plasma filament length is long enough, the impact of the dephasing effect is hard to weaken for changing the THz angular distribution. This is because the concept of point sources \cite{you2012off} ignores the fact that the THz radiation emitting from each source has a broad spectrum and the role of the group velocity $v_{g}$ of the two-color laser pulse is not considered, which makes the THz angular distribution given in the previous models \cite{you2012off} needs to be updated. In our model, we extend the spatial interference of ideal point sources (spherical wave) to the temporal interference of dynamic current sources with a Gaussian temporal profile, determined by the laser pulse duration and the plasma collision, which can present the frequency spectrum of the local radiation emitting from each source. We take the current phase velocity $v_{s}$ as the group velocity of the two-color laser pulse, which can significantly modulate the interference results. With the updated interference and broad spectra of the radiation sources, we can obtain a more self-consistent THz angular distribution and waveforms. Meanwhile, it also provides a way to control the THz angular distribution by adjusting the laser pulse duration and the group velocity to weaken the impact of the dephasing effect.

For fixed filament length $L$, dephasing length $L_{dp}$, and initial relative phase $\phi_{0}$, by changing the current duration $\tau$ and the phase velocity $v_{s}$,we obtain the new angular distribution of THz intensity. Starting from a typical double-peak distribution result with $\tau_{0}$ and $v_{s0}$, an arbitrary coherent term contribution can be described by $\eta=\frac{t^{\prime}}{\tau}=(1-\frac{v_{s0}}{c}\cos\Theta_{0})\frac{\Delta L}{v_{s0}\tau_{0}}$ according to Eq. \textcolor{blue}{(}\ref{eq13}\textcolor{blue}{)}. When the laser group velocity $v_{s0}$ is lowered to $v_{s1}$ or the dynamic current duration $\tau_{0}$ is decreased to $\tau_{1}$, so that:
\begin{equation}\label{eq14}
	\eta= (1-\frac{v_{s0}}{c}\cos\Theta_{0})\frac{\Delta L}{v_{s0}\tau_{0}}=(1-\frac{v_{s1}}{c}\cos\Theta_{1})\frac{\Delta L}{v_{s1}\tau_{1}},
\end{equation}
to cause the THz intensity value at $\Theta_{0}$ shifts towards to a smaller angle $\Theta_{1}$, i.e., double peaks of the angular distribution could shift towards $0^{\circ}$. These phenomenons are displayed in Fig. \ref{fig3}. Note that the current duration $\tau$ can be determined by the laser duration and the laser group velocity $v_{s}$ can be adjusted by the gas density and the laser wavelength. 

Fig. 3(a) shows the change of the THz angular distribution when the current pulse duration is varied. When the current duration is 150 fs (close to the laser duration), the angular distribution has double peaks away from $0^{\circ}$, as usually observed in experiments \cite{zhong2006terahertz}. As the current duration is lessened to 100 fs and 50 fs, the distance between the double peaks is reduced. Further lessening the duration to a very small value of 6 fs, the angular distribution is transited to a single peak at $0^{\circ}$. In Fig. 3(c) we decrease the laser group velocity from $v_{s}=0.9997c$ and find that the THz intensity near $0^{\circ}$ is significantly improved. The dip at $0^{\circ}$ is completely filled with $v_{s}=0.9927c$ and the angular distribution is also transited to a single peak. 

Notice that the evolution of the angular distribution in Figs. 3(a) and 3(c) shows different behaviors. With the decrease of the duration $\tau$, the width of double peaks is reduced and the peaks shift towards the zero angle. While decreasing the velocity $v_{s}$, the width is broadened. This can be understood through Figs. 3(b) and 3(d), in which the angle displacement of the THz intensity can be defined as the difference between the original angle and the new angle. When decreasing the duration $\tau$, the angle displacement is large for a large original angle, causing that the peak width is reduced. While the velocity $v_{s}$ is decreased, the angle displacement is large for a small original angle, giving rise to a broadened peak width. Combined parameter configurations are displayed in Figs. 3(e) and 3(f), these show that the control of the angular distribution can be realized under a more relaxed condition by using the both two methods together.

\begin{figure}[!htbp]
     \centering
     \includegraphics[width=8.6cm]{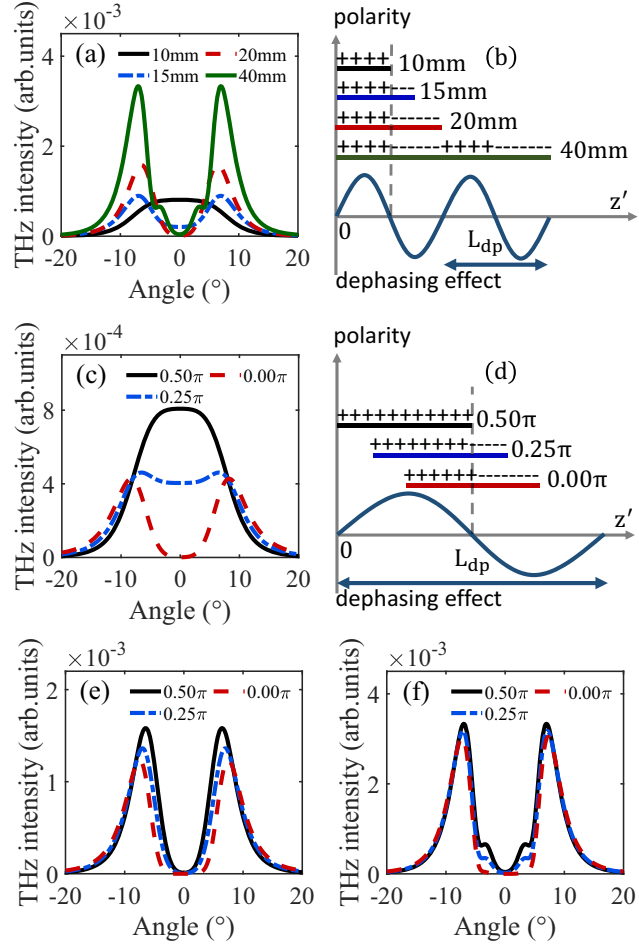}
\begin{spacing}{1.2}	
	\caption{THz angular distributions in different plasma filament lengths and initial phases. (a) Comparison of the angular distribution in different filament lengths $L$ of 5 mm, 10 mm, 20 mm, and 40 mm, where $\phi_{0}=0.50\pi$. A possible interpretation to the dephasing effect is illustrated in (b). (c) Comparison of the angular distribution in different initial phases $\phi_{0}$ of $0.50\pi$, $0.25\pi$ and 0, where $L$=10 mm. A possible interpretion to the dephasing effect is illustrated in (d). [(e), (f)] THz angular distributions in different initial phases of $0.50\pi$, $0.25\pi$ and 0, where we take the filament length $L$=20 mm in (e) and $L$=40 mm in (f). Here, we take $L_{dp}$=20 mm and $\tau$=150 fs.
	\label{fig4}
	}
\end{spacing}
\end{figure}

The relationship among the filament length, the dephasing length and the initial relative phase of the two-color laser field also plays an important role in the angular distribution of THz intensity and the THz yield. For example, the change of the THz yield shows different dependency on the gas pressure in different focus situations \cite{yoo2019highly,rodriguez2010scaling}. In the long-focus condition \cite{yoo2019highly}, the filament length is longer than or comparable with the dephasing length, and increasing the dephasing length by reducing the gas pressure is helpful for improving the THz yield. However, in the short-focus condition \cite{rodriguez2010scaling}, the filament length is much shorter than the dephasing length. Increasing the gas pressure will decrease the dephasing length, but the filament length is too short to be influenced by the dephasing length change. In this case a higher gas pressure causes higher electron density, which can result in a higher THz yield. Another example will be shown below that the dephasing effect can be sufficiently decreased and even avoided by adjusting the initial relative phase for a short filament (e.g., below $L_{dp}$), but for a long filament this method cannot work since the dephasing effect acts periodically.

We change the filament length and the initial relative phase of the two-color laser field and the THz angular distribution is shown in Fig. \ref{fig4}. In Fig. 4(a), as the filament length is increased from 10 mm to 40 mm, the THz angular distribution transits from a single-peak pattern to a double-peak pattern, and the overall THz intensity is enhanced. This result can be explained by Fig. 4(b). Here, we take $L_{dp}$=20 mm and therefore, the polarity of electric field elements for the first 10 mm is positive and it is negative in the following 10 mm.  When $L$=10 mm, all electric field elements have positive polarity, so the THz intensity has a single peak at $0^{\circ}$. When $L$=15 mm, the extra part of 5 mm is negative polarity which makes the THz intensity decrease at $0^{\circ}$, and a small dip appears at this angle. When $L$=20 mm and 40 mm, the number of positive and negative polarity is equal, the dephasing effect at $0^{\circ}$ is enhanced and the dip becomes deeper. Besides, the number of electric field elements is proportional to the filament length, which causes the overall THz intensity is enhanced with the increase of the filament length.

In Fig. 4(c), when changing the initial relative phase from $0.50\pi$ to $0.25\pi$ and 0 (note that the overall THz intensity is very low in this case \cite{wang2013terahertz}), we find that the THz angular distribution transits from a single-peak pattern to a double-peak one. Note that we take $L$=10 mm in this subfigure. The result can be explained by Fig. 4(d). For a given filament length, the polarity of electric field elements can also be changed through adjusting the initial phase. When $\phi_{0}=0.50\pi$, all electric field elements are positive polarity, so the THz intensity has a single peak at $0^{\circ}$. When $\phi_{0}=0.25\pi$, a quarter of the electric field elements become negative polarity, which makes the THz intensity declines at $0^{\circ}$ and a dip appears. When $\phi_{0}=0$, half of the electric field elements become negative polarity, the dephasing effect at $0^{\circ}$ is enhanced and the dip becomes deeper. In Figs. 4(e) and 4(f) we increase the filament length to 20 mm and 40 mm, respectively, and find that the angular distribution of double peaks always appears at any initial phase. All these results suggest that the dephasing effect can influence the THz angular distribution destructively but can be eliminated through adjusting various parameters.

\subsection{THz waveforms from different length filaments}

\begin{figure}[!htbp]
     \centering
    \includegraphics[width=8.6cm]{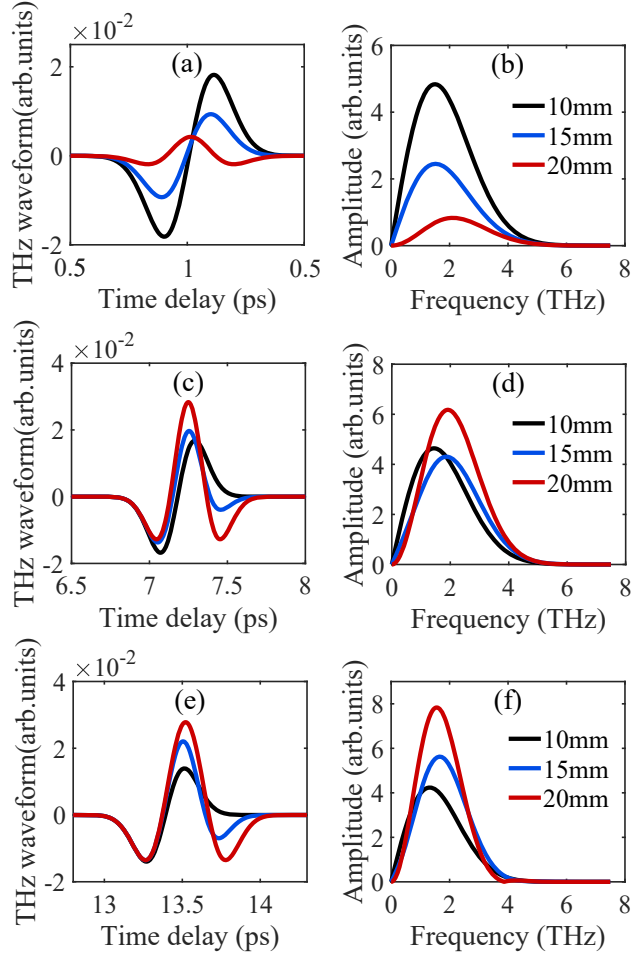}
	\begin{spacing}{1.2}	
		\caption{
         Left column: THz waveforms observed at different angles [(a) $1^{\circ}$, (c) $5^{\circ}$, and (e) $7^{\circ}$]. Right column: the corresponding spectra. In each plot, different curves correspond to the filament lengths of 10 mm, 15 mm and 20 mm, respectively. Here, $\tau=$150 fs, $v_{s}=0.9997c$, $L_{dp}=$20 mm, $\phi_{0}=0.50\pi$ are taken.
			\label{fig5}
		}
	\end{spacing}
\end{figure}

In addition to the angular distribution, the dephasing effect can also influence the THz waveform and the spectrum, as shown in Fig. \ref{fig5}. In this figure we plot THz waveforms and spectra generated from plasma filaments with different lengths and observed at different angles. In Fig. 5(a) the THz pulses are observed at $1^{\circ}$. When the filament length is increased from 10 mm to 20 mm (equal with $L_{dp}$), the waveform is kept as nearly single cycle, but changed from one kind of waveforms observed in experiments \cite{kim2007terahertz,liu2016enhanced} to another in other experiments \cite{cook2000intense,xie2006coherent}. Meanwhile, the THz field strength is decreased due to the dephasing effect. When the observation angle is shifted to $5^{\circ}$ and $7^{\circ}$ [see Figs. 5(c) and 5(e)], the waveforms are similar to the corresponding ones at $1^{\circ}$, but the overall strengths for $L=15$ mm and $L=20$ mm are increased, which are consistent with the angular distribution of THz intensity of double peaks. One can also see that there are obvious time delays between different THz waveforms at different angles since the THz signals are collected at a distance of 500 mm away from the filament and a larger angle causes a longer distance and a longer time delay. Besides, a slight shift of the peak frequency is observed in the THz spectra [Figs. 5(b), 5(d) and 5(f)] with the increase of the filament length, in agreement of the waveform change caused by the dephasing effect.

\subsection{Formation of the THz oscillating tail due to the plasma oscillation}

\begin{figure}[!htbp]
     \centering
     \includegraphics[width=8.6cm]{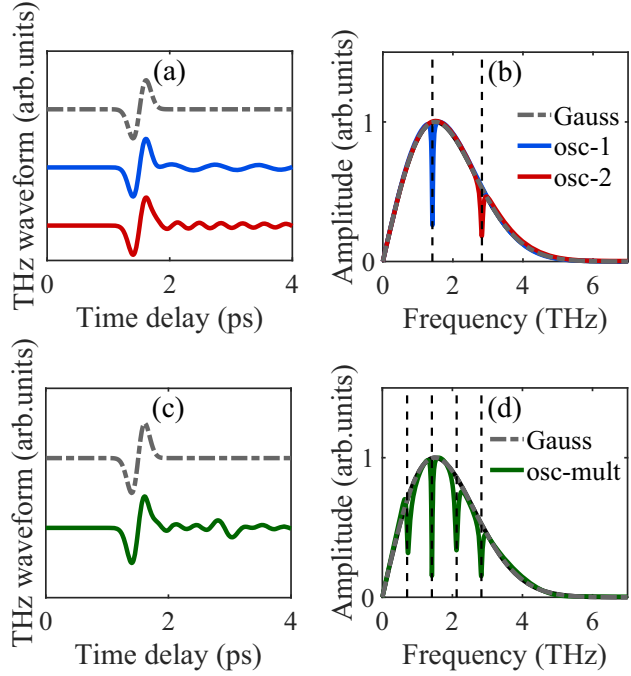}
	\begin{spacing}{1.2}	
		\caption{THz waveforms calculated at $1^{\circ}$ and spectra generated by the plasma currents with different forms. (a) Waveforms generated by the plasma currents with a Gaussian profile (grey dot dash) and a Gaussian pulse plus an oscillating tail with plasma frequencies of 2 THz (blue solid line) and 4 THz (red solid line), respectively. (b) Spectra corresponds to (a). (c) Waveforms generated by the plasma currents with a Gaussian profile (grey dot dash) and a Gaussian pulse plus an oscillating tail containing four plasma frequency components of 1 THz, 2 THz, 3 THz and 4 THz (green solid line). (d) Spectra corresponds to (c). In (b) we plot two vertical lines at $\dfrac{2}{\sqrt{2}}$ THz and $\dfrac{4}{\sqrt{2}}$ THz as well as in (d) four vertical lines at $\dfrac{1}{\sqrt{2}}, \dfrac{2}{\sqrt{2}}, \dfrac{3}{\sqrt{2}}$ and $\dfrac{4}{\sqrt{2}}$ THz.\label{fig6}
		}
	\end{spacing}
\end{figure}

 As described by Eqs. \textcolor{blue}{(}\ref{eq6}\textcolor{blue}{)}-\textcolor{blue}{(}\ref{eq8}\textcolor{blue}{)}, the THz radiation characteristics are mainly determined by the current form $\vec{J}(\vec{x}^{\prime},t)$. In the previous section, we adopt a Gaussian current profile as shown in Eq. \textcolor{blue}{(}\ref{eq9}\textcolor{blue}{)} to contain the finite current duration which can give the single-cycle THz waveform. In this section, we take the current form as a Gaussian current profile plus an oscillating tail to describe the plasma oscillation effect. Accompanied with the current production due to the gas ionization, a plasma is also formed. After the passage of the two-color laser pulse, the electrons gain transverse velocities while the ions are nearly immobile due to much higher mass than electrons. Then, the electrons or currents oscillate transversely around the immobile ions and a periodically-varying charge separation is formed with a plasma oscillation frequency. Amplitude of the current oscillation gradually decreases with the electron collision and the energy loss of THz emission, thus, we take the current form as a Gaussian current profile (the current produced via the ionization) plus an oscillating tail (due to the plasma oscillation).
 
 We consider the charge separation as a small deviation of the electron center from the ion center \cite{zhang2016controllable} as illustrated in Fig. 1(b). The transverse electric field generated by the ion distribution is $E_{R}^{i}=\frac{Zen_{i0}}{2\epsilon_{0}}R$ at $(-r_{0}\le R\le r_{0})$ and the electric field generated by the electron distribution is $E_{R}^{e}=-\frac{en_{e0}}{2\epsilon_{0}}(R-r_{d})$ at $(-r_{0}+r_{d}\le R\le r_0{0}+r_{d})$ under a small displacement $r_{d}$ from the ion center, where $\epsilon_{0}$ is the dielectric constant, $n_{i0}$ is the ion density, and $n_{e0}$ is the electron density. The total electric field is $E=\frac{en_{e0}}{2\epsilon_{0}}r_{d}$ generated by the charge separation with a displacement $r_{d}$ between the ion and the electron center. Inserting this field into the electron motion equation, one can solve an oscillation frequency of the electrons or the current:
\begin{equation}
	\label{eq15}
	\omega_{osc}=\frac{1}{\sqrt{2}}\sqrt{\frac{n_{e0}e^{2}}{m\epsilon_{0}}}=\frac{1}{\sqrt{2}}\omega_{p}.
\end{equation}
This plasma dynamics causes the current oscillates with a frequency $\omega_{osc}$. And thus, the THz pulse also has a frequency component of $\omega_{osc}$.

In Fig. 6(a), we take three different currents: one has a Gaussian profile and the other two have a Gaussian profile plus an oscillating tail caused by the plasma oscillation with plasma frequencies of 2 THz and 4THz separately, the generated THz pulses are shown by the grey dot dash line, blue and red solid lines, respectively. Comparing these three THz pulse waveforms, one can see the plasma oscillation in the current causes a small-amplitude oscillating tail in the THz pulse at a frequency according to Eq. \textcolor{blue}{(}\ref{eq15}\textcolor{blue}{)}. With the plasma oscillation, these THz waveforms are similar to the typical results observed in experiments \cite{tan2022water,kim2007terahertz},which contain a large amplitude near single-cycle waveform lasting for hundreds of femtoseconds and a following weak oscillating tail lasting for several picoseconds. The former can be attributed to the contribution of the Gaussian current and the latter to the plasma oscillation. 

The corresponding spectra are plotted in Fig. 6(b), which show that the plasma oscillation results in sunken spikes around $2/\sqrt{2}$ THz and $4/\sqrt{2}$ THz separately. With the plasma oscillation, the formed sunken spikes in the spectrum are similar to that observed in experiments \cite{liu2015study,kim2007terahertz,liu2016enhanced}. Usually, the oscillating tail in the THz waveform and the corresponding spikes in the spectrum are considered to be caused by the water vapor absorption. It should be pointed out that the experimental comparison with and without vapor absorption shows \cite{liu2010broadband} the spikes in the THz spectrum are very densely distributed in the frequency range of 1 THz to 7 THz. However, the THz spectra observed from two-color laser experiments \cite{liu2016enhanced,kim2007terahertz} have several isolated spikes. Obviously, the plasma oscillation effect could explain such isolated sunken spikes and the corresponding oscillating tail in the THz waveform, as shown in Figs. 6(a) and 6(b). Each isolated spike corresponds to an efficient plasma density component since a filament basically has a non-uniform plasma density profile, which can cause the THz pulse with a few efficient plasma frequencies, as shown in Figs. 6(c) and 6(d). In turn, the spikes in the spectrum could reflect the main plasma density components in the filament. In fact, the plasma oscillation effect can not only result in sunken spikes but also protuberant peaks at corresponding frequencies in the THz spectrum, which is influenced by the time delay between the main Gaussian current and the following oscillating current tail.

\section{Conclusion}
We have developed a far-field model to calculate the angular distribution, the waveform and the spectrum of the THz radiation generated via the two-color laser scheme. We have considered that each field element is originated from a local current source in the plasma filament and the current source has a phase velocity determined by the laser group velocity and a Gaussian temporal profile given by the laser pulse duration and the plasma collision. Then, the spatial interference of ideal point sources in the previous model is extended to the temporal and the spatial interference of dynamic current sources. In this way, the current oscillation caused by the plasma dynamics can be included in our model. The plasma dynamics can result in an oscillating tail in the THz waveform and sunken spikes in the corresponding spectrum, similar to the experimental observation which usually explained by the water vapor absorption. Apart from presenting another possible explanation to the observed oscillating tail, the corresponding spikes in the spectrum could also provide an approach to measure the filament plasma density.  With the current source phase velocity and the temporal profile considered, the laser group velocity and the duration can affect the dephasing effect. Decreasing the laser pulse duration and the group velocity can improve the angular distribution of THz intensity and could even eliminate the dip around $0^\circ$ usually observed in experiments. Adjusting the filament length and the initial relative phase of the two-color laser pulse could vary the angular distribution and the waveform of the THz radiation.

\begin{acknowledgments}
This work was supported by the Strategic Priority Research Program of Chinese Academy of Sciences (Grant No. XDA25050300), the National Key R\&D Program of China (Grant No. 2018YFA0404801), and the Fundamental Research Funds for the Central Universities, the Research Funds of Renmin University of China (20XNLG01).
\end{acknowledgments}

\end{spacing}

\bibliographystyle{apsrev4-2.bst}
%\bibliography{reference}

\end{document}